\documentclass{emulateapj}
\usepackage{graphics,color,amsmath,amssymb,amsfonts}
\newcommand{\ukrts}{ $\mu\mathrm{K}_{\mathrm{\mbox{\tiny\sc cmb}}}\sqrt{\mathrm{s}}$ }

\newcommand{\boomn}{{\sc Boomerang}}
\newcommand{\boom}{{\sc Boomerang} }

\newcommand{\planck}{{\it Planck }}
\newcommand{\planckhfi}{{\it Planck} HFI }

\newcommand{\cmb}{{\sc cmb} }
\newcommand{\cmbn}{{\sc cmb}}

\newcommand{\fwhm}{{\sc fwhm}}

\def\ie{{{\em i.e.}}}

\def\rms{{\em r.m.s.}}

\def\aap{Astr. \& Astroph.\ }

\def\apjl{Astrophys.\ J.\ Lett.\ }
\def\apjs{Astrophys.\ J.\ Suppl.\ }

\def\gev{GeV~c$^{-2}$}

\def\gevpercm3{\gev~\percm3}
\def\gpercm3{g~\percm3}

\newcommand{\plotsize}{0.64}
\newcommand{\halfplotsize}{0.32}

\slugcomment{for submission to The Astrophysical Journal.}
\shorttitle{\boomn03~:~CMB Temperature Power Spectrum}
\shortauthors{JONES ET AL.}

\begin{document}

\title{A Measurement of the Angular Power Spectrum \\ of the CMB
  Temperature Anisotropy from the 2003 Flight of \boomn}
\author{W.~C. Jones\altaffilmark{1},
P.~A.~R. Ade\altaffilmark{2}, 
J.~J. Bock\altaffilmark{3},
J.~R. Bond\altaffilmark{4},
J. Borrill\altaffilmark{5,}\altaffilmark{6}, 
A. Boscaleri\altaffilmark{7},
P. Cabella\altaffilmark{8},\\
C.~R. Contaldi\altaffilmark{4,}\altaffilmark{9}, 
B.~P. Crill\altaffilmark{10},
P. de Bernardis\altaffilmark{11}, 
G. De Gasperis\altaffilmark{8}, 
A. de Oliveira-Costa\altaffilmark{12},\\
G. De Troia\altaffilmark{11}, 
G. Di Stefano\altaffilmark{13},
E. Hivon\altaffilmark{10}, 
A.~H. Jaffe\altaffilmark{9},
T.~S. Kisner\altaffilmark{14,}\altaffilmark{15},
A.~E. Lange\altaffilmark{1},\\
C.~J. MacTavish\altaffilmark{16},
S. Masi\altaffilmark{11}, 
P.~D. Mauskopf\altaffilmark{2}, 
A. Melchiorri\altaffilmark{11,}\altaffilmark{17},
T.~E. Montroy\altaffilmark{15},
P. Natoli\altaffilmark{8,}\altaffilmark{18},\\
C.~B. Netterfield\altaffilmark{16,}\altaffilmark{19}, 
E. Pascale\altaffilmark{16}, 
F. Piacentini\altaffilmark{11},
D. Pogosyan\altaffilmark{4,}\altaffilmark{20},
G. Polenta\altaffilmark{11},
S. Prunet\altaffilmark{21},\\
S. Ricciardi\altaffilmark{11}, 
G. Romeo\altaffilmark{13}, 
J.~E. Ruhl\altaffilmark{15}, 
P. Santini\altaffilmark{11},
M. Tegmark\altaffilmark{12}, 
M. Veneziani\altaffilmark{11},
N. Vittorio\altaffilmark{8,}\altaffilmark{18}}

\altaffiltext{1}{Physics Department, California Institute of
Technology, Pasadena, CA, USA {(wcj@astro.caltech.edu)}}
\altaffiltext{2}{School of Physics and Astronomy, Cardiff University, UK}
\altaffiltext{3}{Jet Propulsion Laboratory, Pasadena, CA, USA}
\altaffiltext{4}{Canadian Institute for Theoretical Astrophysics (CITA),
University of Toronto, ON, Canada}
\altaffiltext{5}{Computational Research Division, LBNL, Berkeley, CA, USA}
\altaffiltext{6}{Space Sciences Laboratory, University of California
  at Berkeley, CA, USA}
\altaffiltext{7}{IFAC-CNR, Firenze, Italy}
\altaffiltext{8}{Dipartimento di Fisica, Universit\`a di Roma ``Tor
  Vergata'', Rome, Italy}
\altaffiltext{9}{Theoretical Physics Group, Imperial College, London, UK}
\altaffiltext{10}{Infrared Processing and Analysis Center, California 
Institute of Technology, Pasadena, CA, USA}
\altaffiltext{11}{Dipartimento di Fisica, Universit\`a di Roma ``La
  Sapienza'', Rome, Italy}
\altaffiltext{12}{Department of Physics, Massachusetts Institute of
Technology, Cambridge, MA, USA}
\altaffiltext{13}{Instituto Nazionale di Geofisica e Vulcanologia,
  Rome, Italy}
\altaffiltext{14}{Department of Physics, University of California at
  Santa Barbara, CA, USA}
\altaffiltext{15}{Department of Physics, Case Western Reserve
University, Cleveland, OH, USA}
\altaffiltext{16}{Department of Physics, University of Toronto, ON, Canada}
\altaffiltext{17}{INFN, Sezione di Roma 1, Rome, Italy}
\altaffiltext{18}{INFN, Sezione di Roma 2, Rome, Italy}
\altaffiltext{19}{Department of Astronomy and Astrophysics, University of 
Toronto, ON, Canada}
\altaffiltext{20}{Department of Physics, University of Alberta,
  Edmonton, AB, Canada}
\altaffiltext{21}{Institut d\rq Astrophysique de Paris, Paris, France}


\begin{abstract}

We report on observations of the Cosmic Microwave Background (\cmbn) 
obtained during the January 2003 flight of \boomn .  These results
are derived from 195 hours of observation with four 145 GHz
Polarization Sensitive Bolometer~(PSB) pairs, identical in design to
the four 143 GHz \planckhfi polarized pixels. The data include 75 hours of
observations distributed over 1.84\% of the sky with an additional 120
hours concentrated on the central portion of the field, itself
representing 0.22\% of the full sky. From these data we derive an
estimate of the angular power spectrum of temperature fluctuations of
the \cmb in 24 bands over the multipole range $50 \leq \ell \leq
1500$.  A series of features, consistent with those expected
from acoustic oscillations in the primordial photon-baryon fluid, are
clearly evident in the power spectrum, as is the exponential damping
of power on scales smaller than the photon mean free path at the epoch
of last scattering ($\ell \gtrsim 900$).   As a consistency check, the
collaboration has performed two fully independent analyses of the time
ordered data, which are found to be in excellent agreement.

\end{abstract}


\section{Introduction}

The wealth of cosmological information that is encoded in the
statistical properties of the Cosmic Microwave Background Radiation (\cmbn)
has motivated a highly successful observational effort to
measure the angular power spectrum of the \cmb temperature anisotropies.
The experimental effort is broad based, with teams reporting results from
interferometric and single dish observations spanning a
decade in frequency and more than three decades in angular
scale~[for some recent results, see~\cite{wmap_bennett,archeops_spectra,retal,dasi_spectra,cbi_mosaic,acbar_spectra,vsa_spectra,maxima_lee}].
The success of these observations, coupled with the predictive
power of accurate theoretical modeling, has contributed to the
ongoing transformation of cosmology into a quantitative, precise, and 
increasingly accurate science.

In this paper, we present the angular power spectrum of the
temperature anisotropies derived from the data obtained during
the January 2003 flight of \boom (hereafter B03).  Having been
upgraded with a polarization sensitive receiver~[\cite{b2kpre,b2k_inst}], B03 
is optimized to probe the polarization of the \cmb at sub-degree
angular scales while retaining full sensitivity to the unpolarized
emission.  In addition to a measurement of the curl-free component of the
\cmb polarization~[\cite{b2k_ee}] and the temperature-polarization
cross correlation~[\cite{b2k_te}], the 2003 flight has resulted in
a precise measurement of the angular power spectrum of the temperature
anisotropies.

These data represent an improvement on published measurements of the 
temperature power spectrum at multipoles $600 \lesssim \ell \lesssim
1200$, which probe physical scales corresponding to the photon mean
free path during the epoch of last scattering.

\section{Instrument and Observations}

Following the successful 1998 Antarctic
campaign~[\cite{nature,netal,retal,b98inst}], an entirely new focal
plane was designed around a set of four 145 GHz Polarization Sensitive
Bolometer (PSB) pairs. These receivers, the first of their kind to be
used in astronomical observations, combine the raw sensitivity of
cryogenic bolometers with intrinsic sensitivity to linear
polarization, a property historically associated only with coherent
detectors~[\cite{wcj_thesis,psbs,tom_thesis}].

In addition to the four 145 GHz PSB pixels, the B03 focal plane
accommodates four dual-frequency photometers, operating in bands 
centered near 245 and 345 GHz.  While the \cmb temperature anisotropies 
are detected with high signal to noise in the 245 and 345 GHz
data~[\cite{tom_thesis,b2k_inst}], these data are not included in the
present analysis.

\begin{figure*}[!t]
\begin{center}
\rotatebox{90}{\scalebox{0.37}{\includegraphics{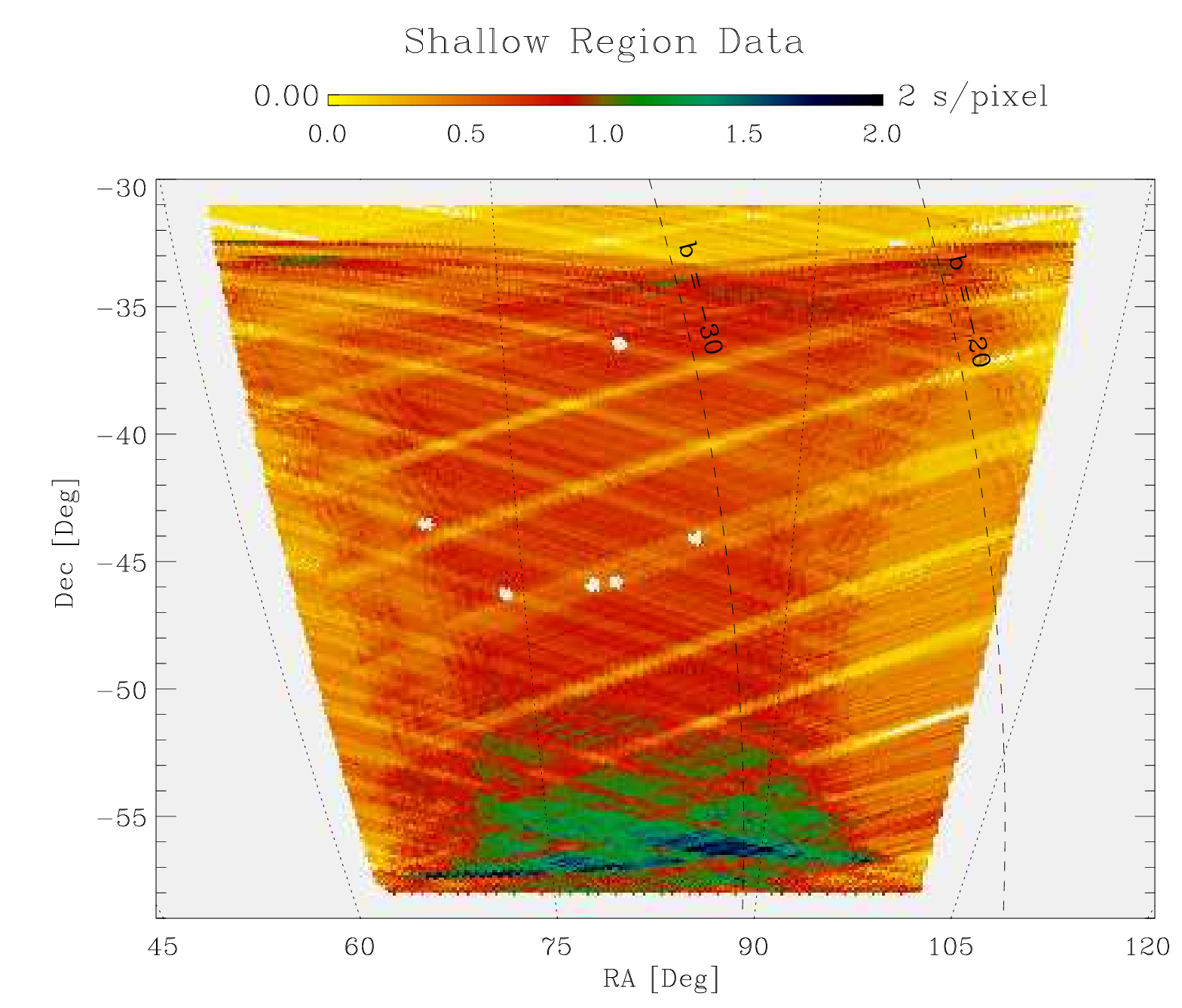}}}\hspace{5mm}
\rotatebox{0}{\scalebox{0.37}{\includegraphics{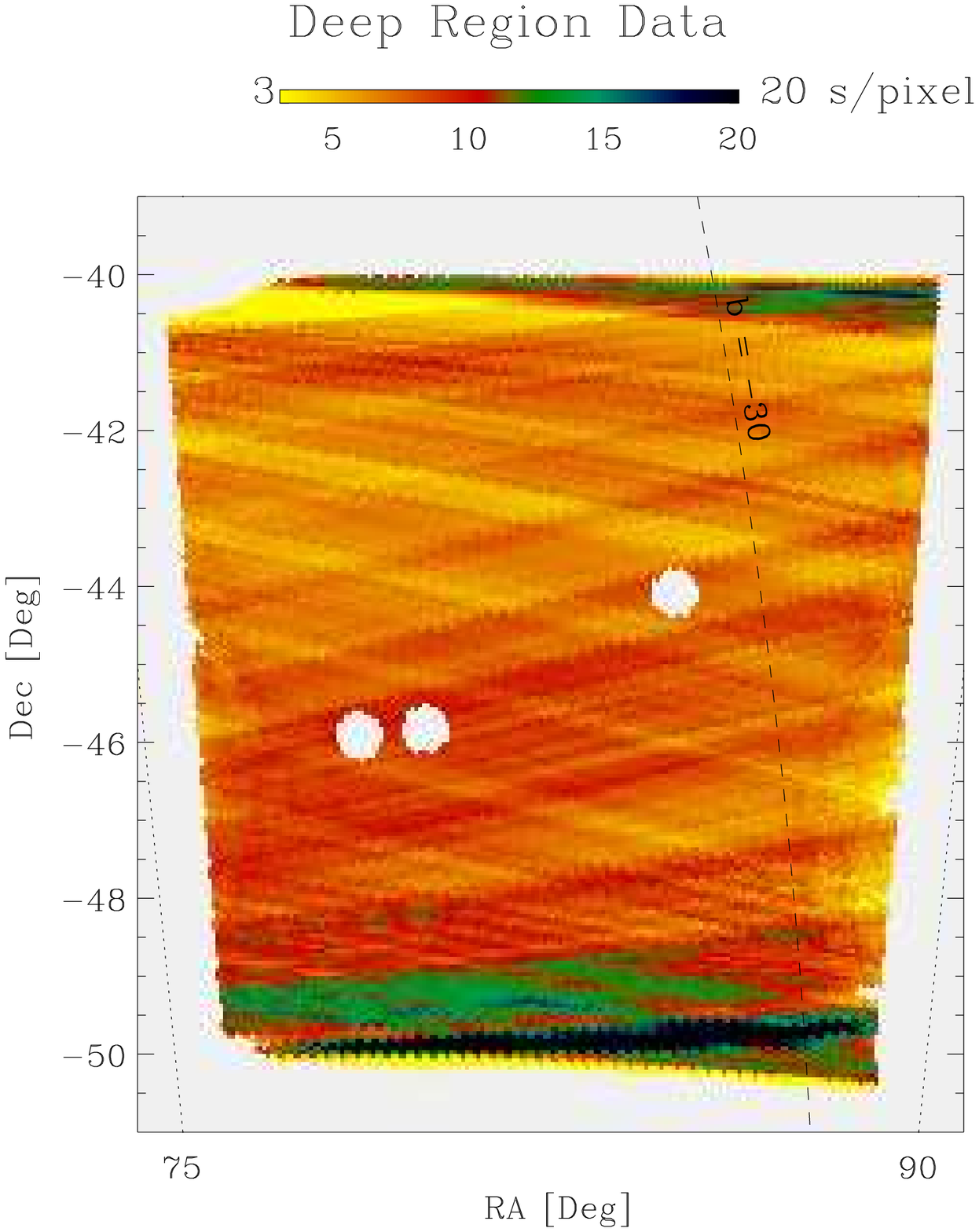}}}
\end{center}
\caption[Sky coverage]{\small The left and right panels show the sky
  coverage obtained during the January 2003 flight of \boomn. The
  integration times per $3.4^\prime$ pixel are shown for the shallow and
  deep data subsets described in the text.  Galactic latitude
  contours of $b=\{-30,-20\}$ are shown for reference. Approximately
  75 hours of observation are devoted to the shallow region, with an
  additional 120 hours concentrated in the central deep field.  The
  cross-linking of the scans due to sky rotation is evident in the
  diagonal striping of the coverage, which has a $\approx 20^\circ$
  opening angle.  Known extragalactic point sources, shown masked in
  these plots, are excluded from the analysis.}
\label{fig:cover}
\end{figure*}

\subsection{Polarization Sensitive \\ Bolometers}

Each bolometer within a PSB pair is sensitive to a linear combination of the
Stokes $I$, $Q$, and $U$ parameters on the sky multiplied by the frequency
response of the receiver, $F_\nu$, and convolved with the
(two-dimensional) co- and cross-polar beam patterns,
$P_\parallel(\hat{r})$ and $P_\perp(\hat{r})$, respectively. After
deconvolving the system transfer function from the time ordered data,
each sample from a single detector within a PSB pair, $d_i$, may be
written as the sum of a signal component
\begin{equation}\begin{array}{l}
d_i=\frac{s}{2}\int d\nu \frac{\lambda^2}{\Omega_b} F_\nu\int\int
d\hat{r}~(P_\parallel(\hat{r}_i)+P_\perp(\hat{r}_i)) \\
\Big[~I+\gamma~\mathcal{P}(\hat{r}_i)~\Big(Q \cos 2\psi_i + U \sin
  2\psi_i~\Big)\Big], \end{array}
\label{eqn:gensig}
\end{equation}
and a noise contribution.  The Stokes parameters are understood to be 
defined on the full sky, and the integration variable is $\hat{r}_i =
\hat{n}_i-\hat{r}$, for a vector, $\hat{n}_i$, describing the pointing
at a time sample, $i$. The normalized beam response and the polarization
efficiency are given by,
\begin{equation}
\begin{array}{ccc}
\mathcal{P}(\hat{r}) \equiv \frac{P_{\parallel}-P_\perp}{P_{\parallel}+P_\perp}
& \hspace{5mm} &
\gamma \equiv \frac{1-\epsilon}{1+\epsilon}
\end{array}.
\label{eqn:tod}
\end{equation}
The polarization leakage parameter, $\epsilon$, is defined as
the ratio of the square of the diagonal elements of the Jones matrix
describing an imperfect polarizer. For B03\rq s PSBs, the leakage
ranges from 5\% to 8\%, as determined from pre-flight measurements.
The angle $\psi$ is the projection of the axis of sensitivity of a
given detector on the sky.  This angle is modulated in time due
to sky rotation, and is further affected by the motion of the gondola.  
The calibration factor, $s$, that converts the brightness
fluctuations in $I$, $Q$, and $U$ to a signal voltage, is obtained
through cross-calibration with the temperature anisotropies observed by
WMAP~[\cite{wmap_general,b2k_inst}].

\begin{deluxetable}{cccccc}
\tabletypesize{\tiny}
\tablefontsize{\small}
\tablecaption{B03 Instrument Summary}
\tablehead{\colhead{$\langle\nu\rangle$}
& \colhead{\underline{{MJy/sr}}}
& \colhead{$\theta_{phys}$} 
& \colhead{$\theta_{eff}{}^a$}
& \colhead{NET${}^b$}
& \colhead{$\sigma_{pix}{}^c$} \\
GHz
& {K$_{\mbox{\sc cmb}}$ }
& \fwhm 
& \fwhm 
& \ukrts
& $\mu$K$_\mathrm{\mbox{\cmb}}$}
\startdata
145 & 388 & 9.95$^\prime$ & 11.5$^\prime$ & 63  & 18\\
245 & 462 & 6.22$^\prime$ & 8.5$^\prime$  & 161 & 50\\
345 & 322 & 6.90$^\prime$ & 9.1$^\prime$  & 233 & 72\\
\enddata
\tablenotetext{a}{The effective beam is defined as the convolution
  of the physical beam with the $\simeq 2.4^\prime$~\rms ~error in the
  pointing reconstruction as determined from point source observations
  in the \cmb field.}
\tablenotetext{b}{The focal plane noise equivalent temperature,
  derived from the inflight noise measured at 1~Hz. Note, however,
  that at both low ($< 100$~mHz) and high ($> 5$~Hz) 
  frequencies, the noise rises significantly.}
\tablenotetext{c}{The approximate noise
  $(\approx\langle\mathrm{diag}(\mathbf{C}_{N})\rangle)$ per $3.4^\prime$
  pixel in the deep field for the data included in this analysis.}
\label{tab:sens}
\end{deluxetable}

A more detailed discussion of the PSB design may be found in~\cite{psbs}. A
description of the characterization of the B03 PSBs may be found
in~\cite{b2k_inst}. A discussion regarding properties and methods of
analysis of PSBs may be found in~\cite{psb_methods} and~\cite{wcj_thesis}.

\subsection{Sky Coverage}

The \boom gondola scans in azimuth, mapping the sky signal to a
bandwidth between $\sim$50 mHz and 5 Hz. Sky rotation modulates
the orientation of the instrument with respect to the signal.
A complete characterization of the Stokes $I$, $Q$ and $U$ parameters is
achieved via a combination of this sky rotation as well as the
joint analysis of the data from all eight detectors in the course of
pixelization. The low frequency stability of the bolometric system and
balloon environment allow an accurate reconstruction of all three
linear Stokes parameters at each pixel with sensitivity to angular
scales ranging from $\sim 10^\circ$ to the $\sim 10^\prime$ beamsize.

The launch date, latitude, and the azimuth/elevation constraints
imposed by solar, limb, and balloon avoidance considerations limit the
region of sky accessible to \boomn.  During the Austral summer, the
anti-solar meridian falls in the vicinity of RA $\simeq 70^\circ$
(J2000 epoch). Continuum emission from the Galaxy is significant over
much of the available sky;  the \cmb field is chosen to minimize this
Galactic contamination, subject to the constraints imposed by the LDB
flight parameters.

The \boom telescope is an Alt-Az mount that is scanned in azimuth at
constant elevation.  The elevation is adjusted on hour time
scales. The pendulation frequencies of the gondola and scan rate
limitations constrain the minimum peak to peak scan amplitude to
approximately $15^\circ$. Earth\rq s rotation provides about
$20^\circ$ of cross linking between scans, which aides in the
decorrelation of the Stokes parameters.  When the local hour angle
(modulo twelve hours) is in the vicinity of the RA of the target field
(\ie, when the sky rotation, and therefore the cross-linking, is
minimal), our scans are redirected to the Galactic plane.

Given B03\rq s instantaneous sensitivity, the amplitude of the \cmb
polarization signal motivates us to concentrate as much integration time as
possible in as small a region as possible.  However, the competing
desire to make high fidelity measurements of (the relatively large
amplitude) unpolarized temperature fluctuations led us to a
two-tiered scan strategy, representing a compromise between
sensitivity to the temperature and polarization power spectra.

During the first four days of the flight, seventy-five hours were
dedicated to observations covering $\simeq 1.84\%$ of the sky (these
observations are referred to as the ``shallow field''), while
an additional 175 hours of observation were concentrated on the
central portion of this field (the ``deep region'').  The deep field
constitutes roughly $\simeq 0.22\%$ of the full sky. The exact
distributions of integration time for the data included in this
analysis are shown in Figure \ref{fig:cover}. The B03 sky coverage is
a subset of the region observed during \boomn\rq s 1998 flight.

\begin{figure}
\begin{center}
\rotatebox{90}{\scalebox{\halfplotsize}{\includegraphics{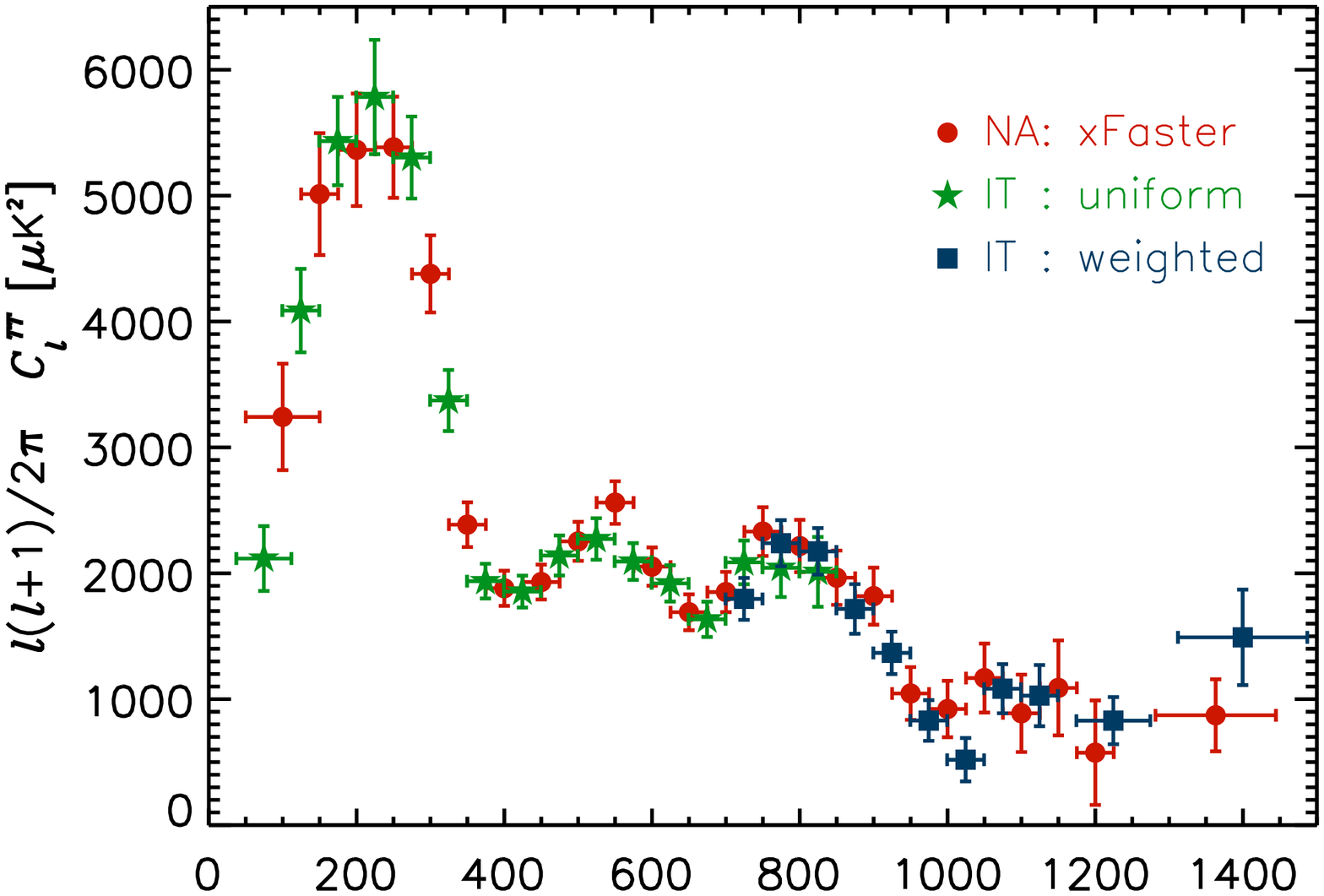}}}
\rotatebox{90}{\scalebox{\halfplotsize}{\includegraphics{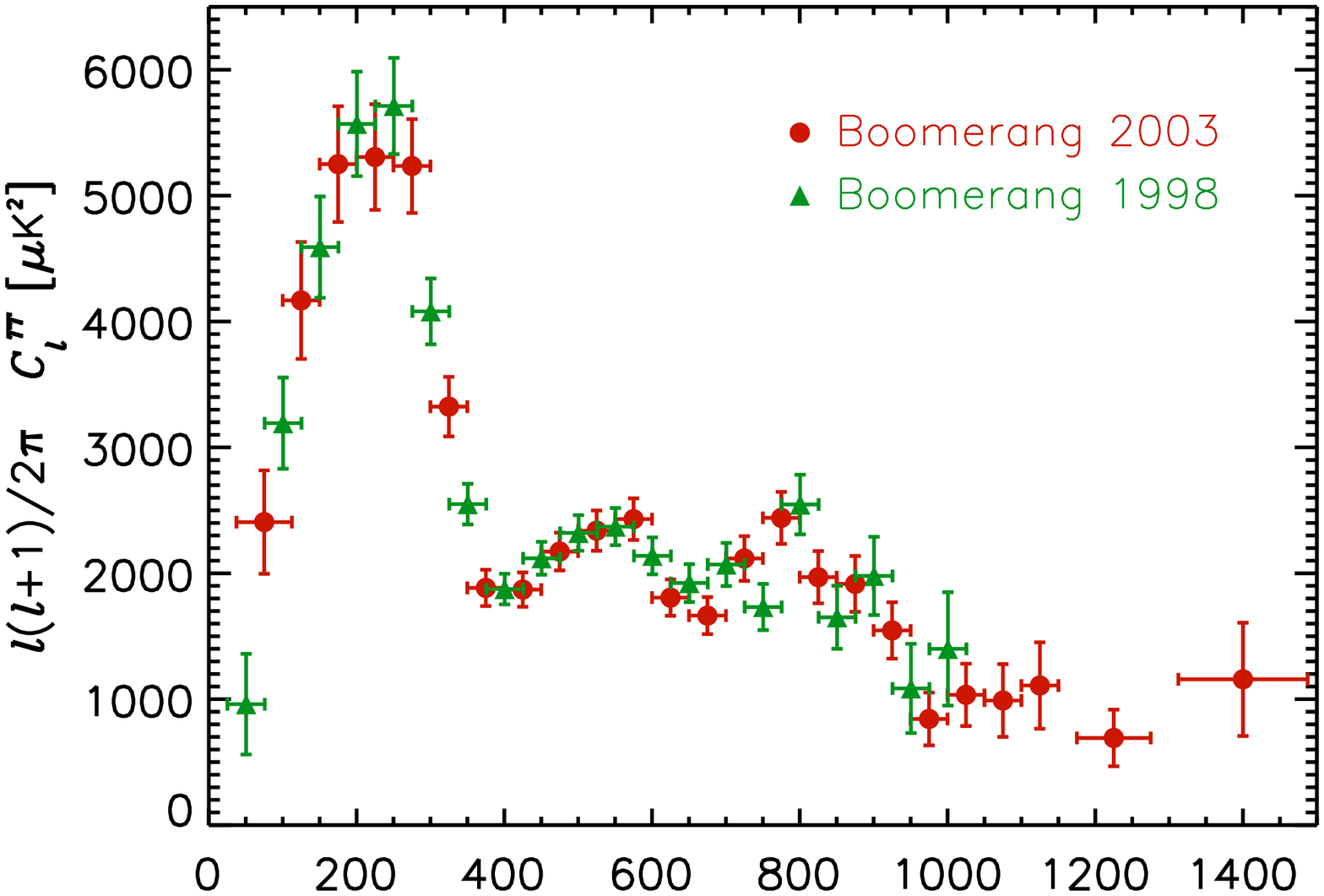}}}
\end{center}
\caption[Comparison of spectra]{\small 
  In the top panel, we show a comparison of the temperature angular power
  spectrum derived from the B03 data using the NA pipeline (xFaster,
  in red circles) and the IT pipeline, the latter using both uniform
  (green stars) and noise weighted (blue squares) masks.  For the IT results,
  uniform weighting of the combined shallow and deep data results in nearly
  optimal errorbars in the signal-dominated regime, while the
  $1/\sqrt{N_{obs}}$ pixel weighting is required to achieve this 
  sensitivity in the noise dominated regime.  The NA spectrum is
  derived using a hybrid auto- and cross- correlation technique which
  achieves nearly optimal errors across the full multipole range.
  In the lower panel, a comparison of the angular power spectrum
  derived from the B98 data~[\cite{retal}], and the B03 results
  reported here.  The 2003 coverage is a sub-set of the region
  observed in 1998, so the signal is completely correlated between the
  two datasets.  The binning of the B03 data in the lower panel is
  shifted by half of a bin relative to the top panel.
}\label{fig:spectrum1}
\end{figure}

\subsection{Inflight Calibration}

Both the calibration of the receivers and the characterization of the
effective beam must be obtained from the inflight data.  The former
is due to the dependence of the bolometer responsivity on the
radiative background, while the latter depends on the fidelity of the
pointing reconstruction.

The calibrations are obtained from the \cmb
itself; both the \cmb dipole and degree-scale anisotropies provide a
well calibrated signal with precisely the desired spectrum.  Scan
synchronous noise limits the accuracy of the dipole calibration to
$\sim 15$\%.  As described in~\cite{b2k_inst}, cross-correlation with
the WMAP data~[\cite{wmap_bennett}] provides an absolute calibration
uncertainty of 1.8\%, including the uncertainty in the WMAP
calibration. The B03 calibration is limited by the relatively low
signal-to-noise of the WMAP data at degree angular scales.  The
relative calibrations of the B03 detectors are obtained from
cross-correlations of single channel temperature maps, and are
accurate to 0.8\%.

The error in the inflight pointing reconstruction is estimated
from the angular size of the extragalactic point sources
in the B03 \cmb fields.  As indicated in Table \ref{tab:sens}, the
effective beams are consistent with the physical optics model of the
optical system, convolved with a $2.4^\prime$ \rms\ Gaussian pointing
jitter.  Measurements of five bright sources in the shallow and deep
fields provide an estimated $0.23^\prime$ \fwhm\ uncertainty on the
width of the effective beams.

\vspace{10mm}\section{Analysis}

As a probe of the robustness of both the data and its analysis, the \boom
team has implemented two fully independent analyses of the B03 time
ordered data. The two pipelines, one centered in North America (NA)
and one in Italy (IT), are described in detail
in~\cite{b2k_inst},~\cite{contaldi},~\cite{psb_methods},~\cite{roma_map} 
and references therein. In this section, we summarize the 
general approach to signal, noise, and angular power spectrum estimation.

\subsection{Signal/Noise Estimation}

Estimates of the signal on the sky and noise in the timestreams are
obtained using an iterative procedure similar to that
applied to the B98 temperature data~[\cite{netal,prunet,retal}].
Least squares maps of the Stokes I, Q, and U parameters are generated from a
combined analysis of the data from all eight 145 GHz PSBs,
corresponding to roughly $3\cdot10^8$ time domain samples.  The data
from each channel are weighted by their NET and combined
during pixelization to decorrelate the linear Stokes parameters
at each pixel. The Stokes parameter maps are generated at $\simeq
3.4^\prime$ resolution, resulting in approximately 377000 spatial
pixels.

\begin{figure*}[!ht]
\begin{center}
\rotatebox{90}{\scalebox{\plotsize}{\includegraphics{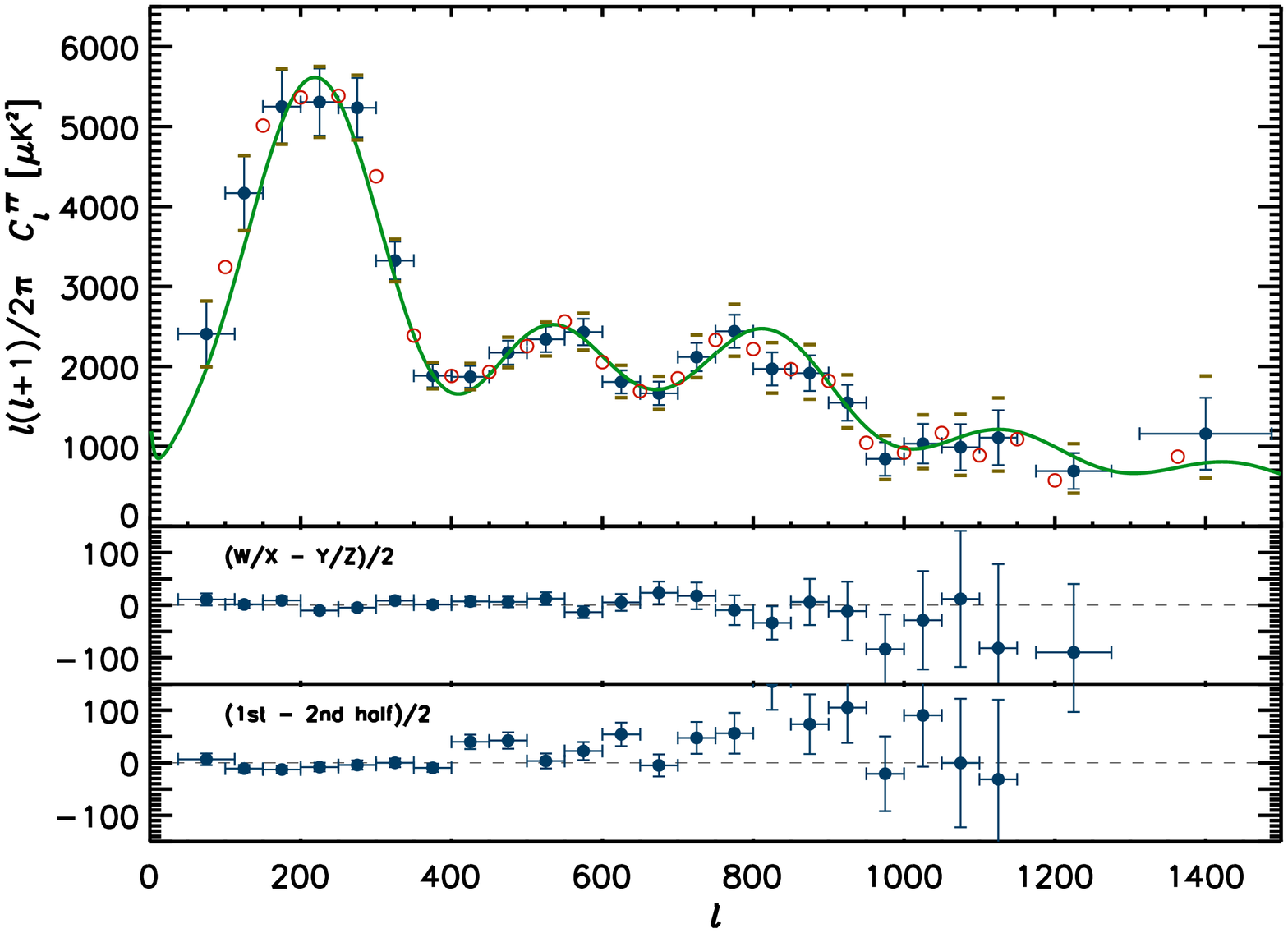}}}
\end{center}
\caption[The B2K $\langle TT\rangle$ spectrum]{\small The temperature
  angular power spectrum, $\mathcal{C}_\ell^{\mbox{\tiny TT}}$, derived from
  the B03 data. The red points represent an alternate binning, and
  should not be interpreted as having statistical weight beyond that of the
  nominally binned data, which are shown with errorbars.  The
  anti-correlations between neighboring bands range from the twelve to
  twenty percent level.  The envelope of the beam uncertainty is
  indicated by ticks bracketing the one sigma errorbars for
  each of the bandpower estimates. As discussed in Section \ref{sec:sys}, these
  limits indicate the amount of ``tilt'' to the spectrum that is allowed by
  the beam uncertainty, and should not be interpreted as an additional
  uncorrelated error in each bin. The solid green line is the concordance
  $\Lambda$CDM model which best fits all published CMB data, including
  the B03 temperature and polarization results.  
  The power spectra of the consistency checks
  described in the text are shown in the lower two panels.
}\label{fig:jack}
\end{figure*}

The NA team divides the timestream into 200 noise-stationary subsets.
The noise power spectra are calculated from each of these chunks using
the converged estimate of the signal derived from the full set of
data.  Each of these spectra are corrected for bias using an ensemble
of signal and noise Monte Carlos.  The IT analysis assumes stationarity of
the noise over the course of the flight.  While the NA approach is more
general, the IT analysis benefits from a reduction in the sample
variance of the noise estimate in the time domain.  Each team employs a
self-consistent treatment of the correlated noise between
channels. Details regarding the two signal and noise estimation
pipelines are provided in~\cite{psb_methods}, and \cite{roma_map}.

\subsection{Power Spectrum Estimation}

Both the NA and IT analysis pipelines employ variations of the
pseudo-${C}_\ell$ Monte Carlo approach to power spectrum estimation 
introduced in~\cite{master}. 
Each incorporates a polarized implementation of a general least
squares signal estimator.  The IT team uses a traditional
pseudo-$C_\ell$ approach using either $1/\sqrt{N_{obs}}$ \emph{or}
uniformly weighted masks to derive the temperature power spectrum,
while a cross-correlation technique similar to that described
in~\cite{tristram}~and \cite{polenta} is used for the polarization analysis.  
Unlike these approaches, the NA estimator of the temperature and
polarization power spectra is based on a hybrid auto- and
cross-correlation technique adapted for multiple data sets
characterized by partially overlapping sky coverage~[\cite{contaldi}].

As described in \cite{master}, the Monte Carlo approach approximates
the spherical harmonic transform of the \emph{heuristically weighted}
data, $\widetilde{C}_\ell^{TT}$, as a convolution over the underlying
power spectrum, $C_{\ell^\prime}^{TT}$, with an additive noise term,
$\widetilde{N}_\ell$. 
\begin{equation}
\widetilde{C}_\ell^{TT} = \sum_{\ell^\prime} K_{\ell\ell^\prime}
B_{\ell^\prime}^2 F_{\ell^\prime} C_{\ell^\prime}^{TT} + \widetilde{N}_\ell
\label{eqn:csky}
\end{equation}

The coupling kernel, $K_{\ell\ell^\prime}$, is derived
analytically from the transform of the weighted mask that has been
applied to the data. The window function, $B_{\ell}$, is the
transform of the effective beam on the sky, and approximates the
convolution over the Stokes parameters appearing in Equation
\ref{eqn:gensig}. The transfer function,
$F_{\ell}$, is a symmetrized approximation of the combined effect of
the scan strategy and time-domain processing on the sensitivity to
a given angular scale.

The transfer function is determined from an
ensemble of signal-only Monte Carlo simulations, while the noise pseudo-power
spectrum, $\widetilde{N}_\ell$, is obtained from an ensemble of
noise-only simulations.  Both the signal and noise simulations
treat the flagged portions of the time stream in exactly the same
manner as the actual data.

\subsection{The B03 Temperature Power Spectrum}

The Monte Carlo approach to power spectrum estimation represents an
approximate treatment of the noise covariance matrix of the map, and
therefore generally results in larger uncertainties than do direct
methods~[\cite{madcap}].  The sensitivity achieved by a Monte Carlo
estimate of the power spectrum is determined by the properties of the noise
and the weighting applied to the data.

\begin{figure}
\begin{center}
\rotatebox{90}{\scalebox{\halfplotsize}{\includegraphics{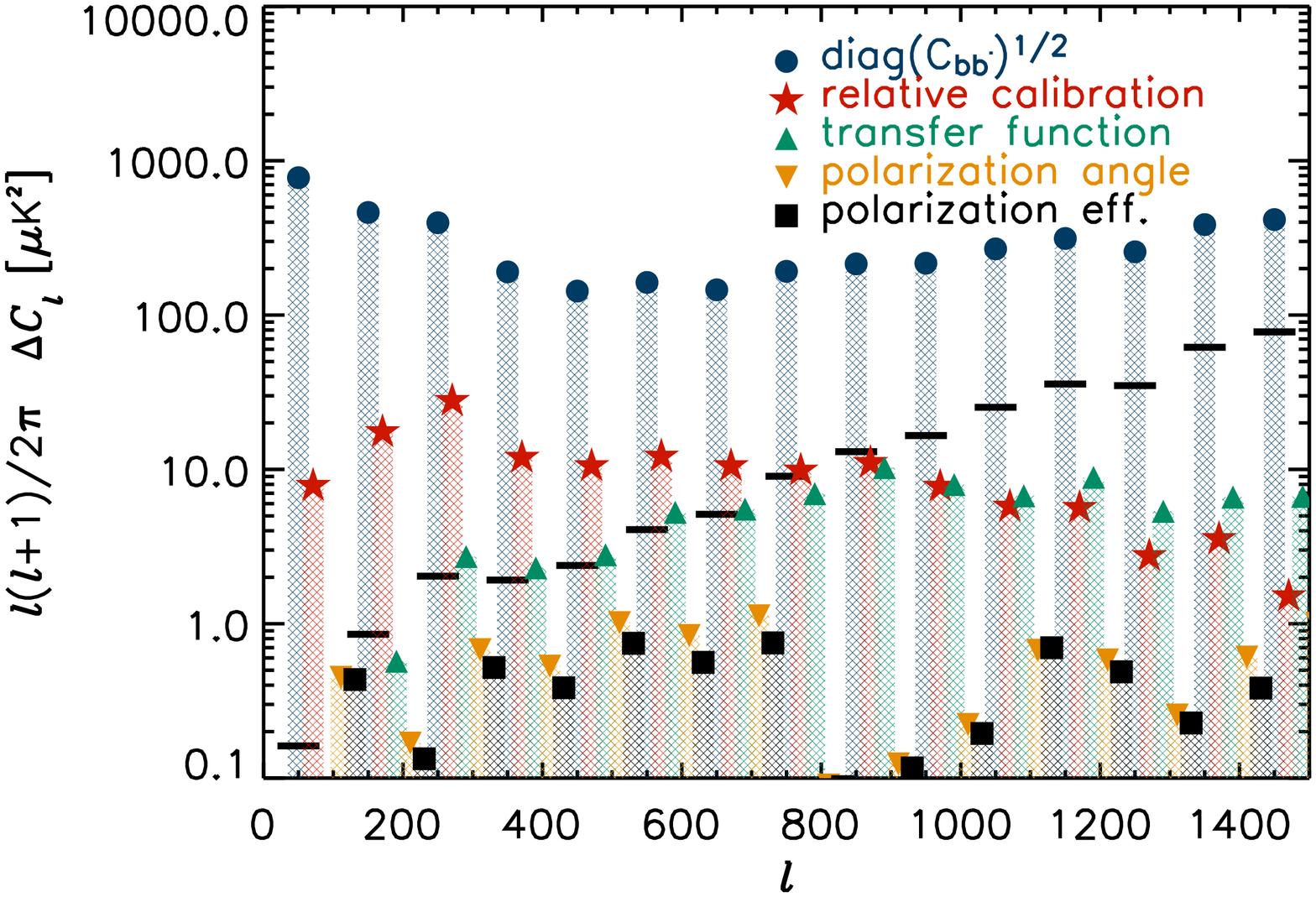}}}
\end{center}
\caption[Systematic error estimates]{\small 
 The propagation of instrumental calibration errors to the bandpower
 estimates.  The amplitude of the statistical uncertainties (sample
 variance plus instrumental noise) for each bin are shown with
 circles. The contributions from the relative calibration, $s \pm0.8$\%, the
 the in-flight transfer function $\tau \pm 10\%$, the polarization
 angle, $\psi \pm 2^\circ$, and polarization efficiency $\epsilon
 \pm 3\%$ are shown for reference.  The horizontal ticks indicate the
 effect of beam uncertainty.  Note that each of these effects are
 highly correlated between bins, and therefore are not properly
 treated by quadrature addition to the statistical uncertainty
 in a given bin. Since the correlation structure of these errors is
 known, they are more properly treated as nuisance parameters, and as
 such are marginalized over during the cosmological parameter
 estimation~[\cite{bridle02}]. For the B03 parameter extraction, only
 the beam error and calibration uncertainty are treated in this
 manner~[\cite{b2k_params}].}
\label{fig:syscheck}
\end{figure}

When analyzing a single map generated from the full set of B03 data,
a complication arises in the selection of the weighting that is
applied to the mask. 
In the limit of white (that is, pixel-uncorrelated) noise, the
optimal power spectrum estimate of a noise-dominated signal is
obtained by weighting the pixels according to $1/\sqrt{N_{obs}}$. For
the combined set of shallow and deep data, this scheme is therefore
virtually equivalent to analyzing \emph{only} the deep region.
Conversely, in the signal dominated regime a uniform weighting
will reduce (to the extent possible) the sample variance in the band
power estimates.  The choice of weighting reflects a trade-off
between the optimality of errorbars at low and high multipoles. 
For the B03 data, the angular power spectrum is
signal dominated at multipoles below $\ell \lesssim 900$, as indicated
in the fourth column of Table \ref{tab:jack}. 

The IT temperature power spectrum is derived from an analysis of the
combined shallow and deep field data, using an approach similar to
that applied to the B98 data in \cite{netal}\ and \cite{retal}.  The IT
analysis is carried out using \emph{either} uniform \emph{or} 
$1/\sqrt{N_{obs}}$ pixel weighting. 
The agreement between the NA and IT analyses, as well as with the B98
results reported in \cite{retal}, is illustrated in Figure \ref{fig:spectrum1}.
These results, representing two independent datasets and no less than three
independent analysis pipelines, are indicative of the robustness of
the result.

The NA pipeline addresses the problem of heuristic weighting of data
with uneven coverage by treating the shallow and deep time ordered
data separately. Each of these subsets, which have roughly uniform
noise per pixel, exhibit partially correlated signal and statistically
independent noise. A uniform weighting is applied to the shallow data,
while the deep data are noise weighted.

Estimates of the underlying power spectra are derived from the joint
analysis of the two auto-spectra \emph{and} the cross-power spectrum
of the shallow and deep data, using a diagonal approximation to the
quadratic Fisher matrix estimator.  The inclusion of both the auto and
cross power spectra allows nearly optimal errors to be obtained over
the full range of angular scales, while computing the complete Fisher
matrix in a self-consistent fashion for all the bandpowers
(temperature and polarization) simultaneously~[\cite{contaldi}].

The full resolution B03 spectrum is displayed with the best fit
concordance $\Lambda$CDM model in the upper panel of Figure
\ref{fig:jack}~[\cite{b2k_params}].  The maximum likelihood
bandpowers, the diagonal components of the correlation matrix, and the
ratio of sample to noise variance are presented in Table
\ref{tab:jack}. As indicated in the fourth column of this table, the
spectrum is sample variance limited at multipoles below $\ell \lesssim
900$. Neighboring bins are anti-correlated at the $\simeq$ 12--20\%
level, the variation resulting from the exact form of the masks and
the multipole binning.

\begin{deluxetable}{rrrrrr}[!ht]
\tabletypesize{\tiny}
\tablefontsize{\small}
\tablecaption{The \boom03 \\ Temperature Power Spectrum}
\tablehead{\colhead{$\ell_b$}
&\colhead{$\mathcal{C}_b$} 
&\colhead{$\Delta\mathcal{C}_b$} 
&\colhead{$\mathcal{C}_b/\mathcal{N}_b$} 
&\colhead{(WX-YZ)/2} 
&\colhead{(1st-2nd)/2}}
\startdata
      75       &    2406       &     410       & 12.28       &      11   $\pm$      11       &       7   $\pm$
      11      \\
     125       &    4167       &     464       & 28.25       &       1   $\pm$       7       &     -11   $\pm$
       7      \\
     175       &    5250       &     459       & 117.66       &       9   $\pm$       7       &     -13   $\pm$
       7      \\
     225       &    5306       &     420       & 132.13       &     -10   $\pm$       5       &      -8   $\pm$
       8      \\
     275       &    5235       &     372       & 125.39       &      -5   $\pm$       6       &      -4   $\pm$
       9      \\
     325       &    3323       &     236       & 72.60       &       9   $\pm$       7       &       0   $\pm$
       8      \\
     375       &    1884       &     144       & 12.89       &       1   $\pm$       7       &     -10   $\pm$
       8      \\
     425       &    1870       &     136       & 7.71       &       7   $\pm$       9       &      40   $\pm$
      14      \\
     475       &    2172       &     149       & 5.97       &       6   $\pm$      10       &      42   $\pm$
      16      \\
     525       &    2338       &     159       & 4.45       &      12   $\pm$      12       &       3   $\pm$
      14      \\
     575       &    2429       &     165       & 3.37      &     -13   $\pm$      11       &      22   $\pm$
      17      \\
     625       &    1806       &     144       & 1.97     &       5   $\pm$      16       &      54   $\pm$
      22      \\
     675       &    1663       &     146       & 1.45      &      24   $\pm$      22       &      -5   $\pm$
      21      \\
     725       &    2117       &     177       & 1.41    &      18   $\pm$      26       &      47   $\pm$
      30      \\
     775       &    2440       &     206       & 1.28      &     -10   $\pm$      28       &      56   $\pm$
      39      \\
     825       &    1968       &     207       &0.90     &     -34   $\pm$      32       &     154   $\pm$
      54      \\
     875       &    1915       &     221       &0.78      &       6   $\pm$      44       &      73   $\pm$
      57      \\
     925       &    1545       &     223       &0.61      &     -11   $\pm$      56       &     105   $\pm$
      67      \\
     975       &     842       &     209       &0.37       &     -84   $\pm$      66       &     -21   $\pm$
      71      \\
    1025       &    1034       &     247       &0.38       &     -29   $\pm$      94       &      90   $\pm$
      98      \\
    1075       &     989       &     289       &0.32      &      12   $\pm$     129       &       0   $\pm$
     122      \\
    1125       &    1108       &     343       &0.32       &     -82   $\pm$     160       &     -31   $\pm$
     151      \\
    1225       &     691       &     224       &0.13       &     -90   $\pm$     130       &     225   $\pm$
     128      \\
    1400       &    1157       &     449       &0.06       &    -611   $\pm$     343       &     702   $\pm$
     301      \\
\enddata
\tablenotetext{}{The angular power spectrum of the B03
  temperature data and the results of the consistency tests described
  in Section \ref{sec:jack}.  The
  bandpowers,~$C_b~\ell_b(\ell_b+1)/2\pi$ and their errors are in units of
  $[\mu\mathrm{K}^2]$. The full set of temperature and
  polarization angular power spectra, and the associated covariance
  matrices and window functions, are publicly available on the
  \boom web servers, \hfill{\tt
    http://cmb.phys.cwru.edu/boomerang/index.html}\hfill and {\tt
    http://oberon.roma1.infn.it/boomerang/b2k}.
  Neighboring bins are anticorrelated at the 12\% to 20\% level,
  depending on the multipole bin.
  The errors,~$\Delta\mathcal{C}_b$, are taken from the diagonal
  elements of the covariance matrix.
  In particular, the diagonal elements of the covariance matrix
  can be approximated by the formula
  $\Delta{C}_b \simeq
  \sqrt{\frac{2}{(2\ell_b+1)~\Delta\ell~f_{sky}~\mathcal{F}_b}} 
  \left(C_b+N_b\right),$
  where $\mathcal{F}_b \leq 1$ is a parameterization of the mode loss
  resulting from the filtering applied to the data. 
  The $N_b$ term represents the (binned and deconvolved)
  power spectrum of the noise, as determined from noise only
  simulations.  Therefore, the ratio $C_b/N_b$ indicates the relative
  contribution of sample and noise variance for a given
  bandpower. This quantity is shown for reference in the fourth
  column.
  The bandpowers obtained from the two consistency tests described in
  \ref{sec:jack} are given in the last two columns.  We have
  subtracted the ensemble averaged power spectra derived from the
  jackknifes of the signal and noise Monte Carlos.  The $\chi^2$ to
  zero for the channel (temporal) jackknife is $\chi^2 = 20.8 (55.5)$
  for 24 degrees of freedom.}
\label{tab:jack}
\end{deluxetable}

The B03 data improve on the published measurements of the
temperature angular power spectrum primarily over the third peak, at
angular scales corresponding to the photon mean free path at the
surface of last scattering.  All of the B03 temperature and
polarization power spectra, inverse Fisher matrices, window
functions and explanatory supplements are publicly available on the
\boom web servers\footnote[1]{{\tt
    http://cmb.phys.cwru.edu/boomerang/index.html} \\ 
    \mbox{\makebox[5.5mm]{}{\tt http://oberon.roma1.infn.it/boomerang/b2k}}}. 

\subsection{Internal Consistency Tests}\label{sec:jack}

As a check on the internal consistency of the data, we perform two complete
jackknife tests in which the TOD are divided in two halves,  and $I$,
$Q$, and $U$ maps are generated from each half independently.  The
power spectra are then computed from the difference of the resultant maps.

The temporal jackknife divides the data into a first half and a
second half, in which the shallow and deep scans are each
divided in order to ensure roughly equal coverage of the sky in each subset.
As discussed in~\cite{b2k_inst} and~\cite{wcj_thesis}, B03 experienced
a dramatic ($\simeq 9$km) loss of altitude over the course of the flight.  
The first-half/second-half test was chosen to provide a check which is
maximally sensitive to systematic effects related to the altitude
drop, such as responsivity drifts, a degradation in the accuracy of
the pointing reconstruction, or atmospheric contamination.

The channel jackknife measures the difference between maps generated
from the two halves of the focal plane.  Each side of the focal plane
accommodates two pairs of PSBs, allowing complete characterization of
the three linear Stokes parameters.  The sky coverage, filtering, and data
flagging of this jackknife test are nearly identical to that of the
full data set.

\begin{figure*}[!ht]
\begin{center}
\rotatebox{90}{\scalebox{\plotsize}{\includegraphics{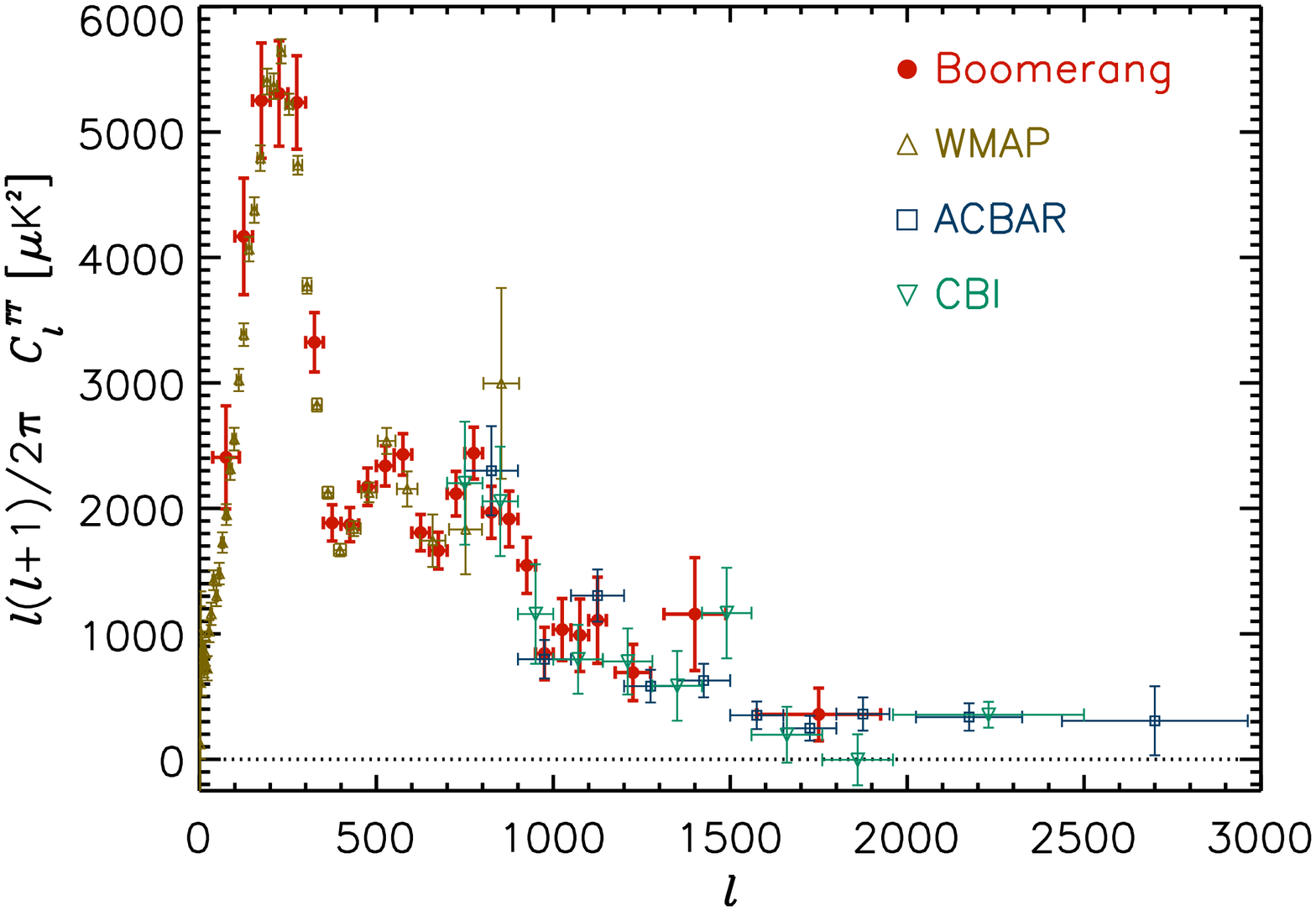}}}
\end{center}
\caption[Compilation plot]{\small A comparison of recently published \cmb
  power spectra, $\mathcal{C}_\ell^{\mbox{\tiny
  TT}}$,~[the data shown are from~\cite{wmap_hinshaw,
  acbar_spectra,cbi_mosaic}, in addition to B03].  The
  data shown are derived from independent measurements that span two orders of
  magnitude in electromagnetic frequency (20-200 GHz), employ both single dish
  and interferometric instruments, and operate from terrestrial,
  balloon-borne, and orbital platforms. The remarkable degree of
  accord between such a diverse set of measurements is indicative of
  the maturity of the observational field.}
\label{fig:comp}
\end{figure*}

The noise properties, and therefore the inverse noise filtering of the
signal, varies slightly from channel to channel. When the data are
processed in the time domain, this asymmetry, combined with the
channel specific flagging of the time ordered data, results in a
non-zero difference spectrum even in a noiseless observation free of
systematic effects.  The temporal jackknife is similarly affected; the
observations are not symmetric for the two subsets of data, resulting
in subtle differences in the signal processing. For both the
channel-based and temporal divisions of the data, this offset is easily
characterized in the Monte Carlo analysis by performing the same
consistency test on an ensemble of simulations. The ensemble average
of these simulated difference spectra are directly subtracted from
those derived from the data.

The high signal to noise ratio with which the temperature anisotropies are
detected is evident in the spectra of the differenced data, which are
shown in the lower two panels of Figure \ref{fig:jack}.
The first of these is the spectrum of the
difference generated from the channel based division, the second is
the spectrum derived from the temporal division.  While the former
(with a $\chi^2 = 20.8$ for 24 degrees of freedom) passes the
jackknife test, the latter clearly fails ($\chi^2
= 55.5$).  The failure of the first-half/second-half consistency test,
though significant, is at an amplitude which is small compared to the
uncertainty from instrumental noise and sample variance in all but the
last two bins, as shown in Table \ref{tab:jack}.

\subsection{Propagation of Systematic Errors}\label{sec:sys}

Uncertainties in the calibration and low level analysis of the data
propagate in a complicated way from the TOD to the band power
estimates.  For the B03 analysis the principle uncertainties are the
determination of the inflight transfer functions, the relative
calibrations, the polarization efficiencies, and the effective beamsize.

A very powerful feature of the Monte Carlo approach is the
ease with which such effects can be modeled \emph{as they appear in
  the time domain}, and properly accounted for in the analysis of the
power spectra.  In particular, we investigate the impact of the
uncertainty in the polarization leakage ($\epsilon$, $\pm 3$\%), the PSB
orientation in the focal plane ($\psi$, $\pm 2^\circ$), and the
relative calibration ($s$, $\pm 0.8$\%) that appear in Equations
\ref{eqn:gensig}~and \ref{eqn:tod}. Additionally, we probe the impact
of the uncertainty on the inflight system transfer function
(parameterized by a single-pole time-constant $\tau$, $\pm 10$\%)
that is used to deconvolve each of the bolometer timestreams at the
earliest stage of the analysis\footnote[2]{The bolometer transfer
  function is dependent on the inflight loading conditions and, for
  B03, is difficult to determine with the required accuracy from the
  inflight data itself. See~\cite{wcj_thesis} and \cite{b2k_inst}
  for further discussion of this topic.}.

A simulated TOD is generated for each B03 channel using a realization
of the best fit concordance $\Lambda$CDM cosmology while randomly varying
the instrumental parameters about their nominal values. An ensemble of
these TODs are processed through the NA analysis pipeline, and the
distribution of the resultant bandpowers are used to propagate
uncertainties on the instrumental parameters to systematic errorbars
on the band power estimates. The results of this analysis, shown in
Figure \ref{fig:syscheck}, indicate that B03 is not limited by
systematics resulting from uncertainties in the characterization of
the instrument.

It is straightforward to analytically determine the impact of an
uncertainty on the effective beamsize on the bandpower estimates.  
An estimate of the beam uncertainty is obtained from fits to the five
brightest extragalactic sources in the \cmb field.  In the limit that the
pointing errors are Gaussian and isotropic, this misestimation will lead
to a bias in spectrum equal to the ratio of the estimated and
underlying window functions.  This bias has a known spectrum
\begin{equation}
{W_\ell^\prime}/{W_\ell} = e^{-\sigma_b^2~(\delta^2+2\delta)~\ell(\ell+1)}~,
\label{eqn:beamerr}
\end{equation}
for an effective beam of width $\sigma_b =
\theta_{\mbox{\scriptsize FWHM}} / \sqrt{8\ln 2}$, and a fractional
error, $\delta$. The envelope of the two-sigma limits
corresponding to a beam uncertainty of $\delta \simeq \pm 2\%$ are shown
as ticks bracketing the errorbars in Figure \ref{fig:jack}.

The effects of both instrumental calibration and beam uncertainty are
highly correlated from bin to bin, and have a known spectrum (Equation
\ref{eqn:beamerr}), or are only weakly dependent on the underlying
spectrum of the signal.  As such, they are both properly treated as
nuisance parameters rather than independent contributions to the
uncertainty in each bin.  In the B03 parameter estimation, the
calibration and beam uncertainty are treated in this manner, following
closely the prescription of~\cite{bridle02}.

\subsection{Foreground Contamination}

Galactic microwave emission can potentially contaminate observations
of the \cmbn.  At 150 GHz, the thermal emission of interstellar dust grains
is expected to be the dominant component of galactic emission
[\cite{brandt94,angelica_fg,masi_dust}]. Therefore any
galactic contamination in the 143 GHz B03 data is expected to exhibit
some spatial correlation with existing infrared surveys.  The spectrum of the
galactic continuum emission is distinct from that of the \cmbn, allowing
multifrequency observations to discriminate between the cosmological
signal and local foregrounds.  For this reason, B03 made simultaneous
observations at 245 and 345 GHz.

Using the method described in~\cite{angelica_dust}, we quantify the
level of foreground contamination in the B03 145 GHz data by
cross-correlating the Stokes I parameter map with the
Schlegel-Finkbeiner-Davis (SFD) dust map. The SFD map is a composite
of the all-sky DIRBE and IRAS 100$\mu$m
surveys~[\cite{dirbe_inst,iras_inst,fink99,schlegel}].

We estimate the uncertainty on the correlation by applying the same
analysis to fifty B03 noise realizations. Before calculating the
correlations, we resample the DIRBE/IRAS data with the B03 scan
pattern and filtering, and convolve both the DIRBE/IRAS data and the
noise with a one-degree Gaussian.  As a consequence, our results are
insensitive to the zero point of the $100\mu$m maps. Using this
procedure, we find no statistically significant correlation between
the 145 GHz B03 and $100\mu$m data in either the B03 deep ($b\lesssim
-30$) or shallow ($b\lesssim -20$) fields from which the \cmb power
spectrum is derived.

The same analysis applied to the 245 GHz and 345 GHz B03 data results
in very clear correlations with the 100 $\mu$m SFD map.  The amplitude of these
correlations, described in more detail in~\cite{b2k_inst}, provides an
empirical scaling of the observed surface brightness of the 100 $\mu$m
dust to 145 GHz. When applied to the SFD map, this empirical scaling implies
\emph{r.m.s} fluctuations due to unpolarized dust of about 1~$\mu$K in the
B03 deep region at the angular scales of interest.  This is roughly
two orders of magnitude below the \cmb contribution, and will
result in a correspondingly negligible impact on the estimate of the
\cmb power spectrum.

\subsection{Features in the Temperature Power Spectrum}

A series of acoustic peaks is readily apparent in the B03 power
spectrum. As has become common practice (see, for example,
~\cite{paolo_peaks,retal,wmap_page}), we calculate a model independent
characterization of the location, amplitude, and significance of the
features in the power spectrum through a comparison of the goodness of
fit of a parabola (or Gaussian) to each set of five contiguous
bins\footnote[3]{While the result is relatively insensitive to the number of
  bins that are fit, as the subsets get larger than the characteristic size of
  the features in the spectrum, the reduced $\chi^2$ clearly will
  degrade.  Five bins is found to be the largest set that does not
  result in a poor goodness of fit statistic over the full range in
  $\ell$.}. In this work, we fit a Gaussian to the band powers in the
vicinity of the first peak, and a parabola to the other features.

As is well known, the likelihoods of the band powers are not Gaussian
distributed~[\cite{bjk98,bjk00}]. We therefore
transform to the ``offset log-normal'' variables whose likelihood
distributions are better approximated by a Gaussian~[\cite{bartlett99,bjk00}].
The transformation is a simple one,
$$Z_b \equiv \ln (\mathcal{C}_b+x_b) $$
where the noise offsets, $x_b$, are determined from the (binned and
deconvolved) $\widetilde{N}_\ell$ as derived from the noise Monte
Carlos.  The inverse Fisher (covariance) matrix must similarly be transformed,
$$(F^Z)^{-1}_{bb^\prime}=\frac{F^{-1}_{bb^\prime}}{(\mathcal{C}_{b}+x_b)(\mathcal{C}_{b^\prime}+x_{b^\prime})}.$$

\begin{figure*}[!h]
\begin{center}
\rotatebox{90}{\scalebox{0.3}{\includegraphics{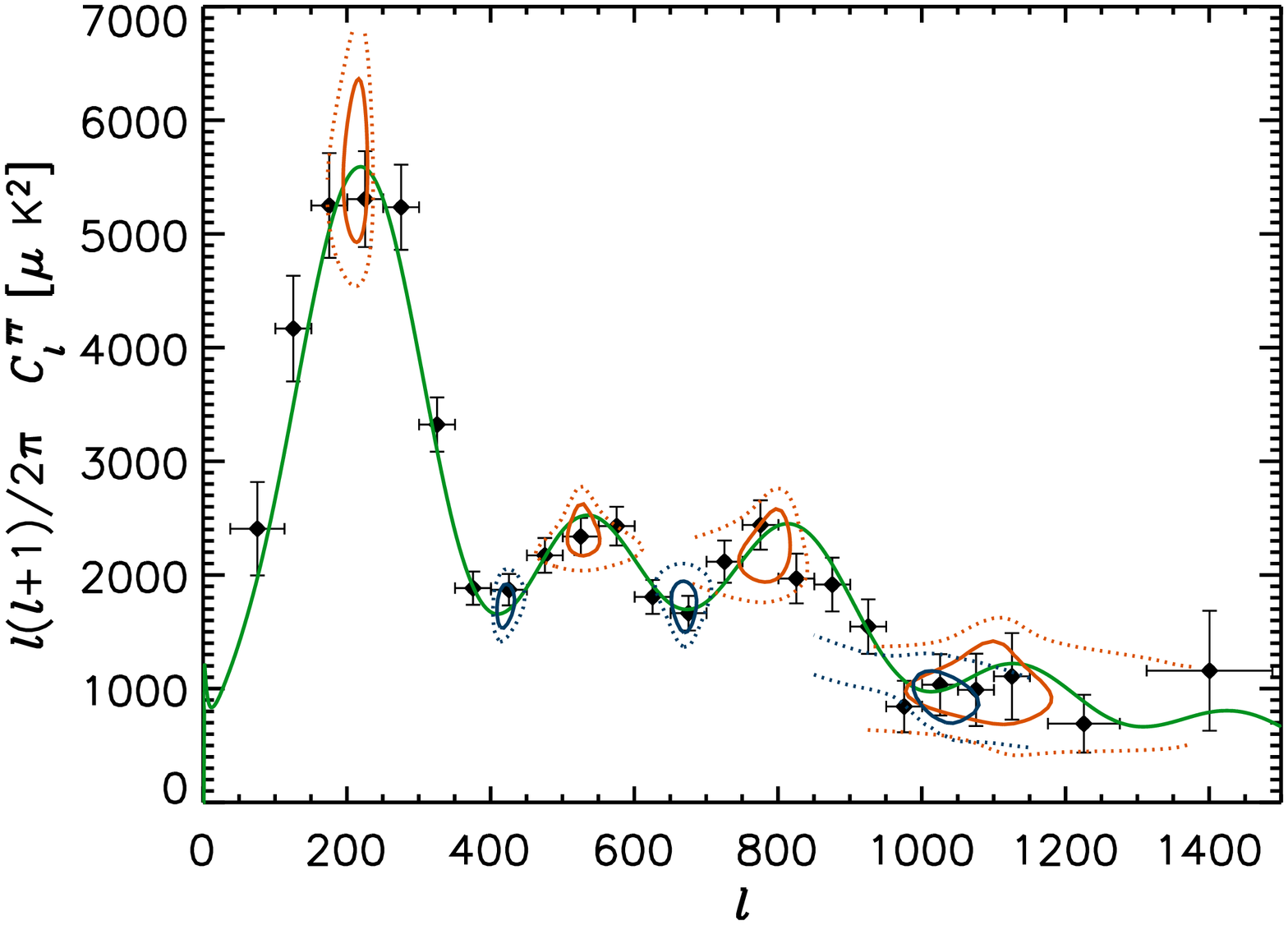}}}
\rotatebox{90}{\scalebox{0.3}{\includegraphics{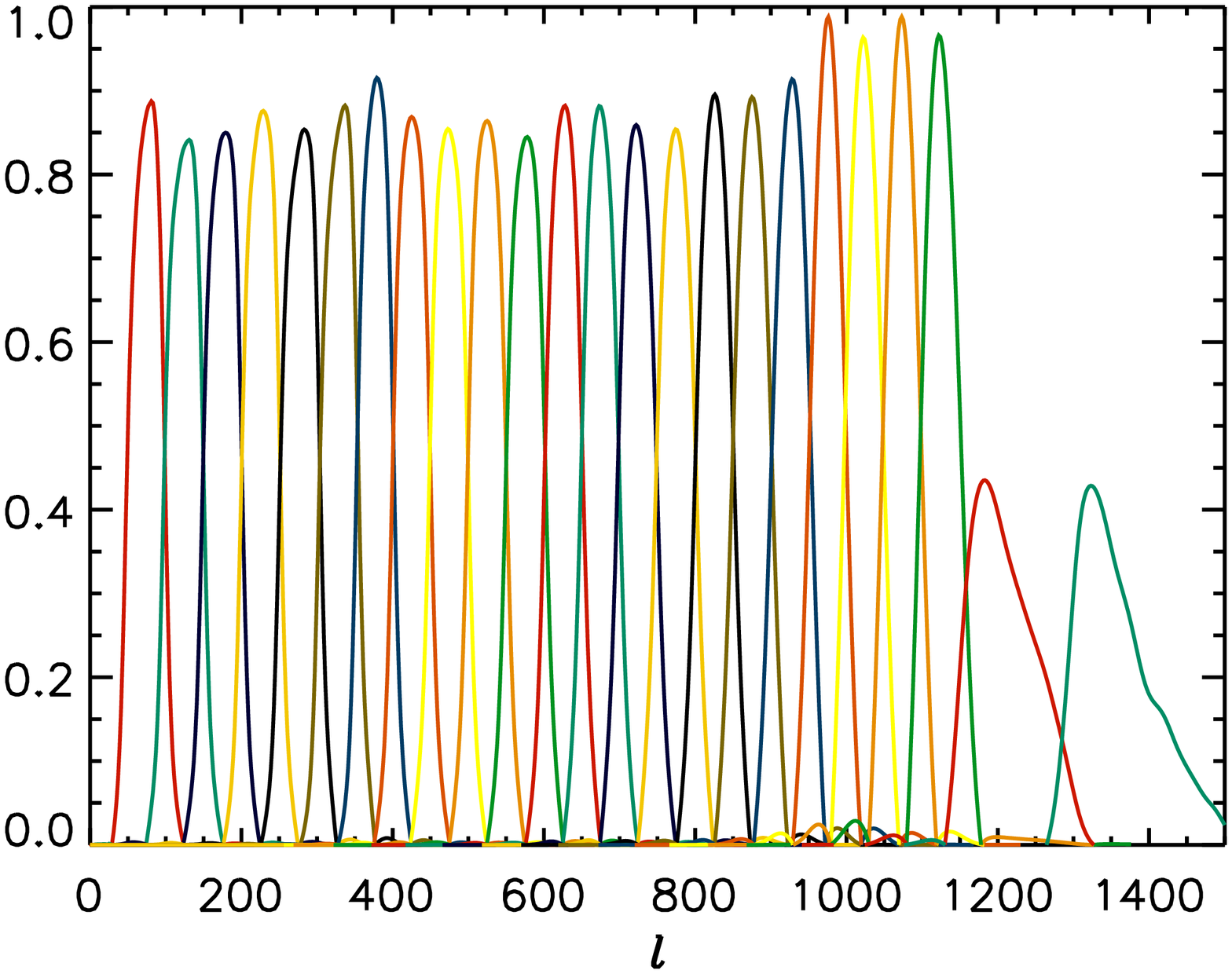}}}
\end{center}
\caption[Features in the $\langle TT\rangle$ spectrum]{\small 
  The left panel shows the \cmb spectrum with the one and two-sigma
  $\Delta\chi^2$ contours for the fits shown in red (blue) for the
  features determined to have negative (positive) curvature.  The
  likelihoods have been marginalized over the curvature parameter.
  The fourth ``peak'' only marginally favors negative curvature over a
  flat bandpower, and is not considered a detection. The right panel
  shows the window functions used in the generation of the band power
  estimates.}
\label{fig:peaks}
\end{figure*}

In addition, the models to be fit are binned into bandpowers
according to the same instrumental window functions, $W_\ell$, 
(shown in Figure \ref{fig:peaks}) which are applied to the raw spectra
$$\mathcal{C}_b^{\mbox{\tiny XX}} \equiv
  \frac{\mathcal{I}\left[W_\ell^b\mathcal{C}_\ell^{\mbox{\tiny XX}}\right]}{\mathcal{I}\left[W_\ell^b\right]}~,$$
where the logarithmic binning operator is defined as
$$\mathcal{I}\left[f_\ell\right]\equiv \sum_\ell
\frac{(\ell+\frac{1}{2})}{\ell(\ell+1)} f_\ell .$$
The parabolic model,
$$\mathcal{C}_\ell^m = \mathcal{C}_c (\ell - \ell_0)^2 +
\mathcal{C}_0$$
is similarly transformed using the offset log-normal approximation, 
and the three dimensional likelihoods are
calculated directly on a grid about each of the best fit
locations. The $\Delta\chi^2$ contours for the curvature-marginalized
likelihoods are shown in Figure \ref{fig:peaks}.

\begin{figure*}[!h]
\begin{center}
\hspace{3mm}\rotatebox{90}{\scalebox{0.3}{\includegraphics{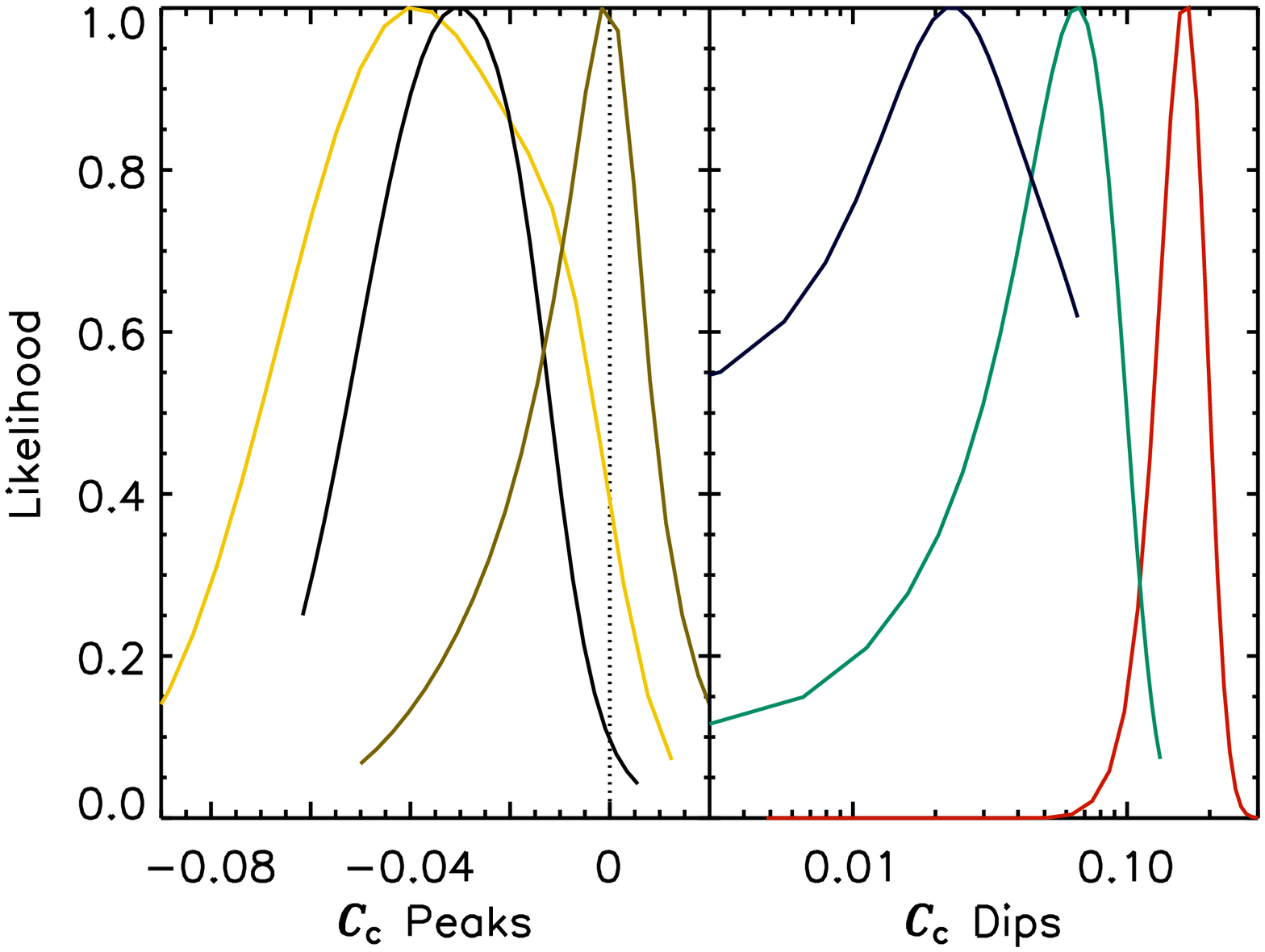}}}
\rotatebox{90}{\scalebox{0.3}{\includegraphics{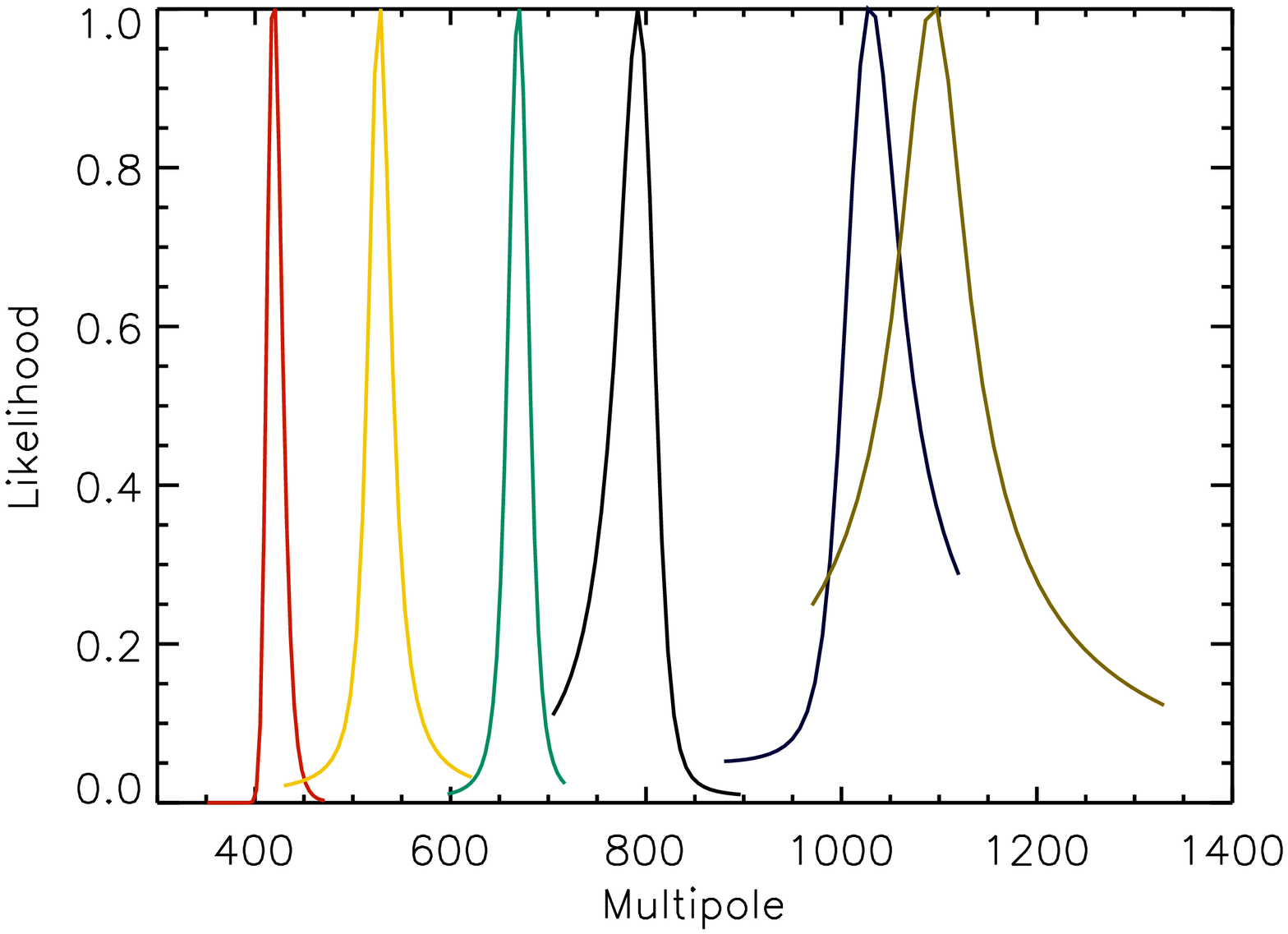}}}
\end{center}
\caption[Peak/Valley likelihoods]{\small The marginalized likelihoods for
the curvature parameter, $\mathcal{C}_c$, (at left) and the multipole,
$\ell_0$, (at right) of each feature in the temperature power
spectrum. All three dips in the power spectrum are detected with high
confidence, whereas only the first three peaks are detected with
curvature significantly different than zero.  The data around the
first peak are fit with a Gaussian rather than a parabolic model, and
the likelihood contours for this feature are not shown.}
\label{fig:likelihood}
\end{figure*}

In previously published B98 data, the significance of a detection
has been determined by the curvature of the likelihood at the peak of the
distribution~[\cite{paolo_peaks,retal}].  However, as is evident in
Figure \ref{fig:likelihood}, the distributions are highly
non-Gaussian. We therefore determine the significance of the
detections from the amplitude of the marginalized likelihood at zero curvature.

The first three peaks and three dips in the power spectrum are
detected with high confidence. Although the data favor a fourth peak
in the vicinity of $\ell = 1055$, with an amplitude of
$\mathcal{C}_{1055} = 1020~\mu K^2$, the marginalized likelihood for
the curvature parameter is not inconsistent with zero.  For
comparison, the same analysis was applied to the B98 data from the
\cite{retal} release, as well as  the binned first-year WMAP
data~[\cite{wmap_bennett}]. The results of all three analyses are
compared in Table \ref{tab:peaks}, and indicate a remarkable degree of
consistency between the three independent experiments.

The degree of concordance in temperature observations is further
illustrated in Figure \ref{fig:comp}, which shows a compilation of
the power spectrum estimates derived from four experiments,
representing four very different experimental approaches, with
observations probing nearly a decade in electromagnetic frequency.

\section{Conclusion}

We derive an estimate of the angular power spectrum of the \cmb from
the data obtained with the 145 GHz polarization sensitive bolometers
that flew on the January 2003 flight of \boomn. 
The 245 and 345 GHz channels place stringent limits on the level of
foreground contamination. We characterize the systematic effects that
result from various instrumental calibration uncertainties with Monte Carlo
analyses, and verify the consistency of the data with jackknife tests.

The B03 data are the first to be obtained using PSBs, which are identical
in design to the polarized pixels in the \planckhfi focal plane. The
sensitivity per resolution element achieved in the B03 deep field is
comparable to that anticipated by \planck at 143 GHz.
The high signal to noise of the temperature data results in a sample
variance limited estimate of the power spectrum at multipoles $\ell
\lesssim 900$.

The B03 data presented in this work represent the most
precise measurements to date of the angular power spectrum between
$600 \lesssim \ell \lesssim 1200$.  In this regard, B03 plays a
valuable role in bridging the gap between the all-sky WMAP survey at
$\ell \lesssim 600$, and the high angular resolution data from
terrestrial observations above $\ell \gtrsim 1200$. We characterize a
series of features in the power spectrum which extend to multipoles
$\ell \gtrsim 1000$, consistent with those expected from acoustic
oscillations in the primordial plasma in the context of standard cosmologies.

\section*{Acknowledgements}

We gratefully acknowledge support from the CIAR, CSA, and NSERC in Canada, ASI,
University La Sapienza and PNRA in Italy, PPARC and the Leverhulme
Trust in the UK, and NASA (awards NAG5-9251 and NAG5-12723) and NSF
(awards OPP-9980654 and OPP-0407592) in the USA. Additional support
for detector development was provided by CIT and JPL. CBN acknowledges
support from a Sloan Foundation Fellowship, WCJ and TEM were partially
supported by NASA GSRP Fellowships. Field, logistical, and flight
support were supplied by USAP and NSBF; data recovery was particularly
appreciated. This research used resources at NERSC, supported
by the DOE under Contract No. DE-AC03-76SF00098, and the MacKenzie
cluster at CITA, funded by the Canada Foundation for Innovation. We
also thank the CASPUR (Rome-ITALY) computational facilities and the
Applied Cluster Computing Technologies Group at the Jet Propulsion
Laboratory for computing time and technical support. Some of the
results in this paper have been derived using the HEALPix
package~[\cite{healpix}], and nearly all have benefitted from the FFTW
implementation of discrete Fourier transforms~[\cite{fftw}].  

\begin{deluxetable*}{lllllll}[!ht]
\tabletypesize{\scriptsize}
\tablefontsize{\small}
\tablecaption{Features in the Temperature Power Spectrum}
\tablehead{\colhead{feature} 
&\colhead{$\ell_\mathrm{B03}$} 
&\colhead{$\Delta T_\mathrm{B03}^2$} 
&\colhead{$\ell_\mathrm{B98}$}
&\colhead{$\Delta T_\mathrm{B98}^2$} 
&\colhead{$\ell_\mathrm{WMAP}$} 
&\colhead{$\Delta T_\mathrm{WMAP}^2$}}
\startdata
Peak 1 & 
214 {\tiny $\begin{array}{c}+9 \\ -12\end{array}$} &
5614{\tiny $\begin{array}{c}+450 \\ -443\end{array}$} & 
217 {\tiny $\begin{array}{c}+10 \\ -10\end{array}$} &
5551{\tiny $\begin{array}{c}+477 \\ -443\end{array}$} & 
222 {\tiny $\begin{array}{c}+3 \\ -2\end{array}$} &
5385{\tiny $\begin{array}{c}+147 \\ -157\end{array}$} \\
Valley 1 & 
413 {\tiny $\begin{array}{c}+10 \\ -5\end{array}$} &
1717{\tiny $\begin{array}{c}+133 \\ -70\end{array}$} & 
411 {\tiny $\begin{array}{c}+9 \\ -7\end{array}$} &
1870{\tiny $\begin{array}{c}+136 \\ -120\end{array}$} & 
418 {\tiny $\begin{array}{c}+5 \\ -4\end{array}$} &
1660{\tiny $\begin{array}{c}+62 \\ -62\end{array}$} \\
Peak 2 & 
529 {\tiny $\begin{array}{c}+14 \\ -6\end{array}$} &
2419{\tiny $\begin{array}{c}+125 \\ -128\end{array}$} & 
526 {\tiny $\begin{array}{c}+17 \\ -14\end{array}$} &
2316{\tiny $\begin{array}{c}+119 \\ -121\end{array}$} & 
530 {\tiny $\begin{array}{c}+15 \\ -8\end{array}$} &
2404{\tiny $\begin{array}{c}+89 \\ -64\end{array}$} \\
Valley 2 & 
659 {\tiny $\begin{array}{c}+12 \\ -11\end{array}$} &
1780{\tiny $\begin{array}{c}+131 \\ -165\end{array}$} & 
(677) {\tiny $\begin{array}{c}+65 \\ -29\end{array}$} &
(1958){\tiny $\begin{array}{c}+200 \\ -170\end{array}$} & 
--- &
--- \\
Peak 3 & 
781 {\tiny $\begin{array}{c}+15 \\ -22\end{array}$} &
2166{\tiny $\begin{array}{c}+208 \\ -216\end{array}$} & 
(766) {\tiny $\begin{array}{c}+42 \\ -43\end{array}$} &
(2080){\tiny $\begin{array}{c}+261 \\ -227\end{array}$} & 
--- &
--- \\
Valley 3 & 
1015 {\tiny $\begin{array}{c}+26 \\ -23\end{array}$} &
991  {\tiny $\begin{array}{c}+137 \\ -192\end{array}$} & 
---&
---& 
---&
--- \\
Peak 4 & 
(1055) {\tiny $\begin{array}{c}+58 \\ -56\end{array}$} &
(1024) {\tiny $\begin{array}{c}+254 \\ -271\end{array}$} & 
--- &
--- &
--- &
---
\enddata
\tablenotetext{}{A
  comparison of the locations and amplitudes of the features in the
  temperature power spectrum derived from the B03, B98, and
  WMAP datasets.  The values and ($1\sigma$) errors are obtained from the
  marginalized likelihood distributions directly.  Values in parenthesis
  indicate a curvature parameter consistent with zero at $2\sigma$
  level, or greater (that is, a marginalized likelihood for the
  curvature parameter for which $\mathcal{L}(\mathcal{C}_c=0)\geq
  2\sigma$). Note that this analysis has been performed on the binned
  WMAP data for better comparison with the \boom window functions.  
  The full resolution WMAP spectrum provides constraints on the first
  peak and dip locations that are stronger than (but also consistent
  with) those reported here.  For the full resolution analysis of the
  first year WMAP data, see~\cite{wmap_page}.}
\label{tab:peaks}
\end{deluxetable*}

\clearpage

\bibliographystyle{apj}
\bibliography{tt_short}

\end{document}